\documentclass{aa}
\usepackage{graphicx,svg,bm,amsmath,dsfont,siunitx,wasysym,tikz,color,scalerel,multirow,tabularx,placeins}
\usepackage{natbib,twoopt}
\usepackage[breaklinks=true]{hyperref} 
\bibpunct{(}{)}{;}{a}{}{,}             
\makeatletter
  \newcommandtwoopt{\citeads}[3][][]{\href{http://adsabs.harvard.edu/abs/#3}%
    {\def\hyper@linkstart##1##2{}%
     \let\hyper@linkend\@empty\citealp[#1][#2]{#3}}}
  \newcommandtwoopt{\citepads}[3][][]{\href{http://adsabs.harvard.edu/abs/#3}%
    {\def\hyper@linkstart##1##2{}%
     \let\hyper@linkend\@empty\citep[#1][#2]{#3}}}
  \newcommandtwoopt{\citetads}[3][][]{\href{http://adsabs.harvard.edu/abs/#3}%
    {\def\hyper@linkstart##1##2{}%
     \let\hyper@linkend\@empty\citet[#1][#2]{#3}}}
  \newcommandtwoopt{\citeyearads}[3][][]%
    {\href{http://adsabs.harvard.edu/abs/#3}
    {\def\hyper@linkstart##1##2{}%
     \let\hyper@linkend\@empty\citeyear[#1][#2]{#3}}}
\makeatother
\definecolor{midblue}{rgb}{0.0,0.4,0.7}
\definecolor{midgreen}{rgb}{0.0,0.8,0.3}
\definecolor{mypurple}{rgb}{0.8,0.2,0.8}
\definecolor{midorange}{rgb}{0.8,0.4,0.0}

\newcommand{\eqnl}[2]{\begin{eqnarray}\label{#1}#2\end{eqnarray}}

\newcommand{\intef}[4]{\int \limits_{#1 = #2}^{#3} \! #4 \, \mathrm{d} #1}

\newcommand{\ave}[1]{\langle #1 \rangle}

\newcommand{\dd}[0]{\mathrm{d}}
\definecolor{midpurple}{rgb}{0.7,0.1,0.7}
\newcommand{\s}[2]{{#1}_{\mathrm{#2}}}
\newcommand{\RA}{{\mathsf{RA}}}
\newcommand{\Dec}{{\mathsf{Dec}}}

\newcommand{\norm}[1]{\lVert #1 \rVert}

\begin{document}

\title{Almost fifty years of Metsähovi solar observations on 37 GHz with recovered digitised historical maps.}
\titlerunning{Fifty years of Metsähovi solar observations on 37 GHz}
\author{
%
%
  Sami Kivistö \inst{1,2}\orcid{0000-0002-7693-6468}
  \and
  Frederick Gent\inst{2,3,5,6,7}\orcid{0000-0002-1331-226}
  \and
  Merja Tornikoski \inst{1}\orcid{0000-0003-1249-6026}
  \and
  Joni Tammi\inst{1}\orcid{0000-0002-9164-2695}
  \and\\
  Maarit Korpi-Lagg\inst{2,3,4}\orcid{0000-0002-9614-2200}
  }

\institute{
  Aalto University Mets\"ahovi Radio Observatory (MRO), Mets\"ahovintie 114, 02540 Kylm\"al\"a, Finland
  \and
  Department of Computer Science, Aalto University, PO Box 15400, 00076, Finland\\
       \email{sami.k.kivisto@aalto.fi}
  \and
  Nordic Institute for Theoretical Physics, Roslagstullsbacken 23, 106 91 Stockholm, Sweden
  \and
  Max Planck Institute for Solar System Research, Justus-von-Liebig-Weg 3, 37077 G\"ottingen, Germany
  \and
  Stockholm University, SE-106 91 Stockholm, Sweden
  \and
  KTH Royal Institute of Technology, SE-100 44 Stockholm, Sweden
  \and
  School of Mathematics, Statistics and Physics, Newcastle University, NE1 7RU, UK
  }
\date{Received \today; accepted TBA}


  \abstract
  {
    Aalto University Mets\"ahovi Radio Observatory has collected solar intensity maps for over $45$ years. Most data coverage is on the $\SI{37}{GHz}$ frequency band, tracking emissions primarily at the chromosphere and coronal transition region. The data spans four sunspot cycles or two solar magnetic cycles.
  }
  {
    We present solar maps, including recently restored data prior to 1989, spanning 1978 to 2020 after correcting for observational and temporal bias.
  }
  {

    The solar maps consist of radio intensity sampled along scanlines of the
antenna sweep. We fit a circular disk to the set of intensity samples,
neglecting any exceptional features in the fitting process to improve
accuracy. Applying a simple astronomical model of Sun and Earth, we assign
each radio specimen its heliographic coordinates at the time of
observation.  We bin the sample data by time and heliographic latitude to
construct a diagram analoguous to the classic butterfly diagram of sunspot activity.
  }
  {
    Radio butterfly diagram at $\SI{37}{GHz}$, spanning
    solar cycles 21 to 24 and extending near to the poles.
  }
  {
    We have developed a method for compensating for seasonal and atmospheric bias
    in the radio data, as well as correcting for the effects of limb brightening and
    beamwidth convolution to isolate physical features. Our observations are consistent with
    observations in nearby bandwidths and indicate the possibility of polar cyclic behaviour
    with a period exceeding the solar 11 year cycle.
  }

  \keywords{Solar physics, radioastronomy, butterfly diagram, solar cycle}

  \maketitle

\section{Introduction}

  The solar magnetic field is known to switch polarity in a full cycle of
approximately 22 years, which consists of two sunspot cycles
(\citeads{1919ApJ....49..153H}, \citeads{1961ApJ...133..572B}).  At the beginning of each sunspot cycle, the solar
magnetic field is an axial dipole.  Solar differential rotation transforms the
predominantly axisymmetric poloidal magnetic field into an azimuthal field with
opposite polarity between hemispheres.  In the presence of convection and the
Coriolis force due to solar rotation, this azimuthal field is further distorted
by the action of turbulent vortical flows stretching, twisting and looping the
field, known as the $\alpha$-effect (\citeads{1966ZNatA..21..369S}, \citeads{1980mfmd.book.....K}).  After approximately
11 years the large scale magnetic field evolves into a dipolar field with
polarity reversed from its original sign and the next sunspot cycle begins.
The full 22 year cycle comprises two such sunspot cycles.

  Starting from a quiet Sun containing a relatively weak axial dipole, each
sunspot cycle emerges as activity on the visible surface of the Sun, typically
near mid-latitudes ($\pm20^{\circ}$ to $\pm 30^{\circ}$).  While the frequency
and coverage of the sunspots increases towards a maximum, the location of the
activity drifts equatorward.  This migration continues after the maximum with
decreasing activity closer to the equator, bringing the sunspot cycle to an
end.  This was first recorded in the Maunder butterfly diagram by Edward
Maunder in 1904.  Since sunspots are a feature of Sun's magnetic field, and as
the solar magnetic field is mostly axisymmetric, the longitudinal location of
the activity can reasonably be neglected \citepads{2006A&A...460..875P}.  Considering this
two-dimensional time-latitude histogram of sunspot activity or magnetic field
intensity, similar butterfly diagrams are also recovered from solar
magnetograms (\citeads{1983ApJ...265.1056G}, \citeads{2012ApJ...749...27V}, \citeads{2017A&A...599A.131L}).  It is also reported for
radio brightenings of the 17 and $\SI{34}{GHz}$ band
(\citeads{2013PASJ...65S..17S}, \citeads{2014ApJ...790..134S}) and of $\SI{37}{GHz}$ \citepads{2018AN....339..204K}.

  From sunspots and magnetograms of the photosphere we have relatively rich
well-resolved data about the two-dimensional (2D) evolution of the solar
surface magnetic field, but observation of the three-dimensional (3D) structure
of the field is limited by the weakness of magnetic measurements from the solar
atmosphere relative to the surface.  Mets\"ahovi Radio Observatory
(MRO)\footnote{
  \href{https://www.aalto.fi/en/metsahovi-radio-observatory}{
        https://www.aalto.fi/en/metsahovi-radio-observatory}}
  is observing emissions, primarily in $\SI{37}{GHz}$ frequency band, dating
back over four decades to the late 1970s.  This is in a frequency range suitable
for detecting non-thermal emissions from high energy electrons associated with
active-region magnetic fields, so is a good proxy for the intensity of the
magnetic field. A detailed analysis by \citetads{2020SoPh..295..105K} relate brightening at
$\SI{37}{GHz}$ emissions with coronal holes and magnetic bright points in
extreme UV emissions.  These are also well suited to observing the thermal
emissions from the quiet Sun.  Comparing and contrasting the maps from these
emissions with other observations can help to reveal more about the structure
of the magnetic field in plasma of varying temperature and by inference height.
Furthermore, the data incorporate almost four magnetic cycles, yielding a
valuable resource to investigate long-term variation within and between cycles,
alongside and in contrast to other sources.  Similar bands at $\SI{950}{GHz}$
and $\SI{84}{GHz}$ have probed the chromesphere \citepads{2019ApJ...877L..26L}, but
without the long term historical record available here.

Previous works \citepads[][]{2018AN....339..204K} have studied the statistics of active
regions, radio samples with excess of a fixed threshold, sampling individual
MRO solar maps of the cycle 24.  Analysis based on manually collected values
has been made, such as \citetads{2012BaltA..21..255K}, \citetads{2012AN....333...20K} and \citetads{1987A&A...184...57V}.  Cycle 21 maps have
previously been analysed with manual methods in \citetads{2012BaltA..21..255K} and \citetads{2012AN....333...20K} using
undigitised maps.

For the first time, the whole 45 years of data are available in a form that is
fully accessible to automation, allowing detailed analysis and in accurate
heliographic coordinates \href{http://urn.fi/urn:nbn:fi:att:f371cb6d-f84c-4d76-99e4-c39c639fd0de}{\citep{mrosolardata}}. The earliest of these maps
feature the solar cycle 21 and have not been available until appropriate
digital processing tools were implemented in \href{https://urn.fi/URN:NBN:fi:aalto-201809034908}{\citet{masterthesis}}.  The new
analysis includes, along with further development of the calibration process,
calculating temporal averages of radio samples within a given region on the
rotating heliographic surface.  Heliographic features can be distinguished
based on their actual separateness.

  Across the lifetime of the MRO the equipment has experienced several upgrades
and evolution of the data format.  This article shall describe the pipeline of
data processing to transform the raw data into a time series of solar radio
maps, normalised, corrected for instrumental bias, atmospheric noise and
effects, time of day and seasonal effects and solar limb dimming effects.  We
present two methods for correctly detecting the position of the solar disk in
each map and three methods for normalising the signal levels, such that the
background \emph{black sky} has value zero and the Quiet Sun Level (hereafter
QSL), the solar intensity in the absence of surface magnetic activity, has
value unity.  The maps are then projected from Right Ascension (RA) and
Declination (Dec) onto their rotating heliographic surface coordinates. This is described in
Sect.~\ref{sect:methods}. We
we have also prepared tools for automatic tracking of active regions,
which are suitable for any magnetograms or similar maps, and which we shall
apply in detail in future publication.

  We present a time-latitude histogram of the radio intensity, including
previously inaccessible machine-drawn maps from 1978 -- 1987 and up to 2024.
This yields the most complete time series thus far available from MRO, spanning
almost completely the last four sunspot cycles.  We present our butterfly
diagrams  in Sect.~\ref{sect:results} and discuss in Sect.~\ref{sect:discussion}
how MRO observations
compare with the results of other observations, including \citetads{2013PASJ...65S..17S}.


  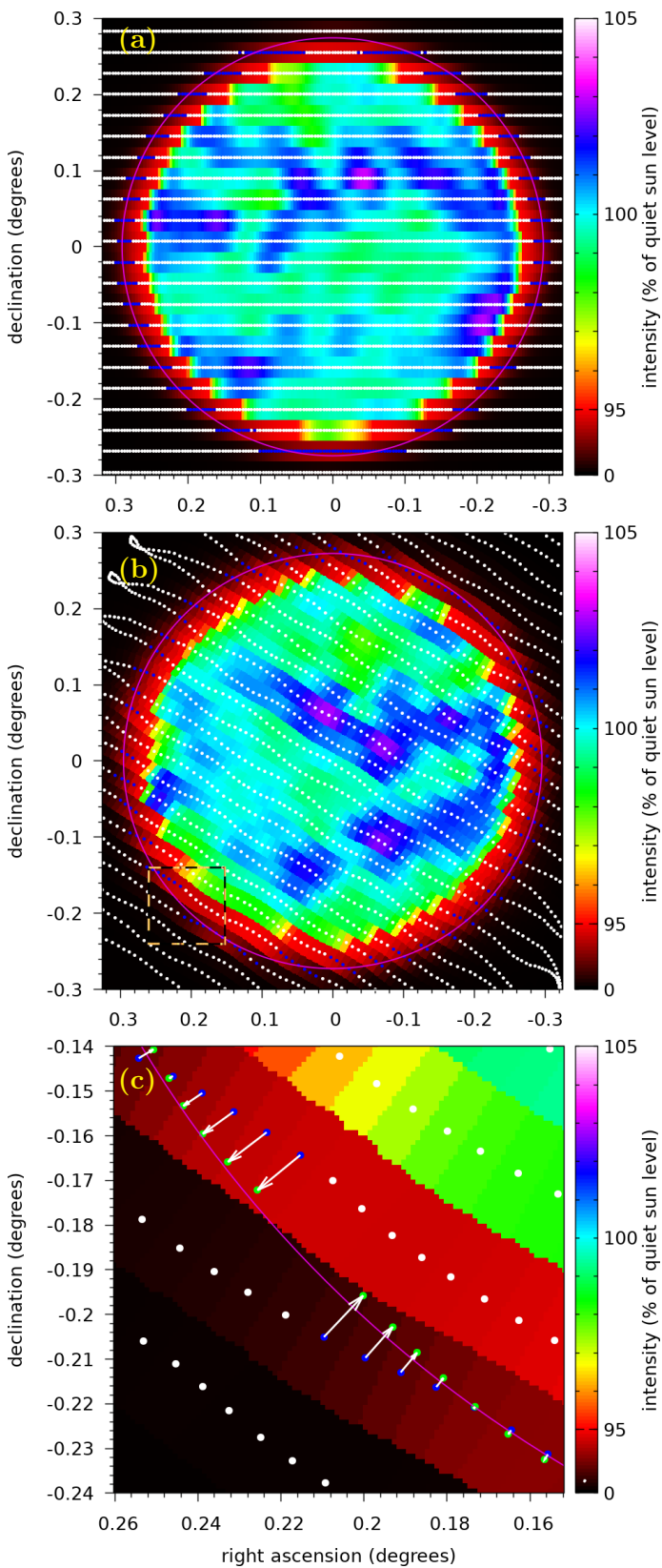
\begin{figure}
  \centering
  \includegraphics[width=\columnwidth]{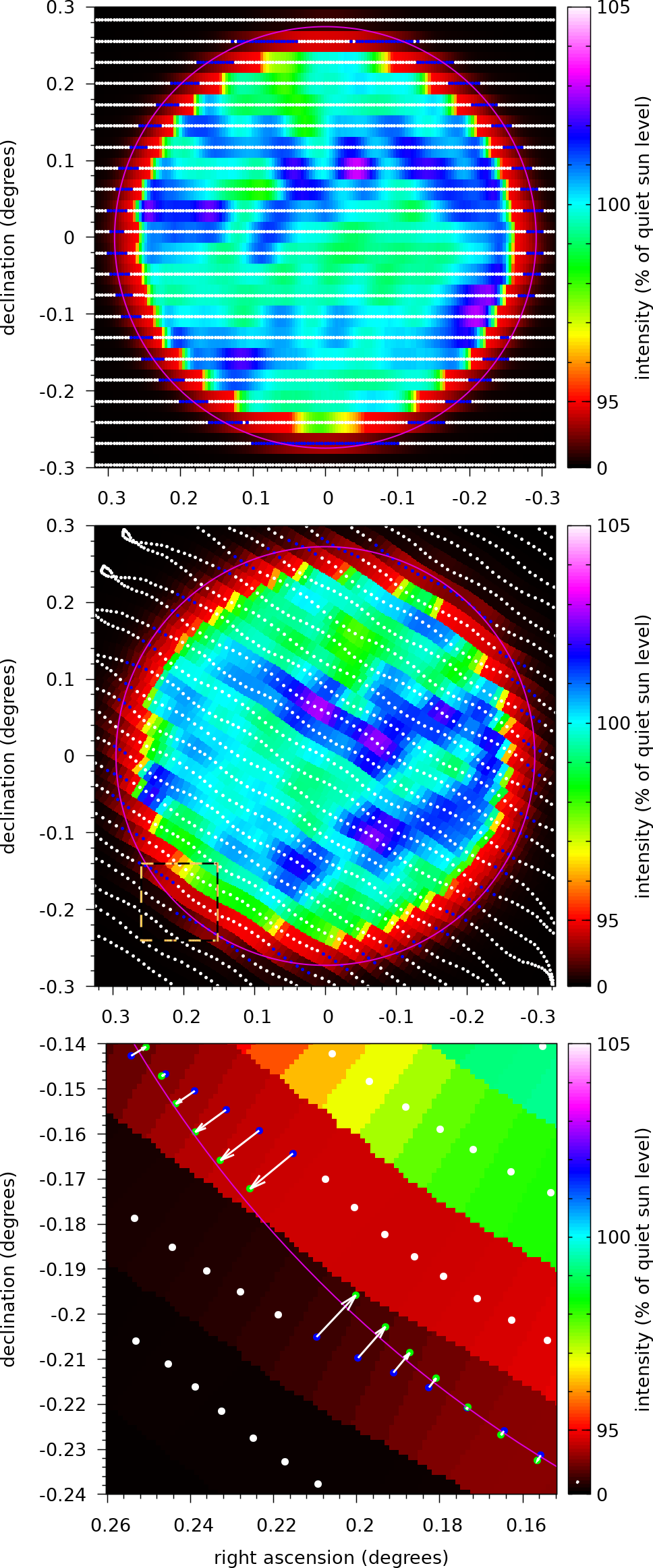}
  \begin{picture}(0,0)(0,0)
    \put(-85,612){\begin{large}\color{yellow}{\bf{(a)}}\end{large}}
    \put(-85,402){\begin{large}\color{yellow}{\bf{(b)}}\end{large}}
    \put(-85,200){\begin{large}\color{yellow}{\bf{(c)}}\end{large}}
  \end{picture}
  \caption{
    Typical Mets\"ahovi solar observations on $\SI{37}{GHz}$ {(a)}
    25.07.2011 and {(b)} 05.07.2016. Prior to May 2015 the observation beam
    scans along the equator, with samples mapped onto a rectangular grid.
    Since June 2015 a more accurate fit to the beam path was used, scanning
    horizontally. For the intensity color map each pixel is mapped to the
    nearest observed sample.
    This is an original uncalibrated map.
    {(c)} Closeup of the 05.07.2016 map which
    illustrates circle fitting.
    \label{oldmap}\label{typicalmap}}
  \end{figure}

\section{Methods}\label{sect:methods}

Here, we detail the techniques applied to collate Metsähovi observations up to present,
spanning four solar cycles.  The data include multiple formats, cadence and
observational conditions, so require normalisation to compare results over
time.  To accurately relate observed features to the solar surface, the maps
must be centered and the chromospheric radius of the observations accurately
determined.  The telescope has measurement biases and the measured intensities
are affected by the altitude of the observation or location on the solar disk.
We apply corrections to better identify features near the limb.

\subsection{The solar maps}\label{sect:source}

During a typical measurement, the antenna beam makes a sweep across the visible
solar disk (Fig.~\ref{typicalmap}).  The historical record comprises three
major convention sets applying to the data.

  \begin{enumerate}[A]
    \item
    Since June 2015 the sweeps have been horizontal, with shifting altitude
    between sweeps, as shown in Fig.~\ref{typicalmap}{(b)}.
    These modern maps respect the true path of the antenna, typically recorded
    with $25$ samples per second for two minutes.

    \item
    Until May 2015 the maps were scanned in the equatorial direction,
    so that declination is shifted between adjacent sweeps, as displayed in
    Fig.~\ref{typicalmap}{(a)}.
    The maps assume linear antenna tracks with constant speed and sample rate.

    \item
    Solar maps from 1978 to 1987 were originally recorded on magnetic
    tapes, from which the data were rendered onto paper contour maps using a
    mechanical plotter.
    Subsequently the magnetic tapes were lost, so that only the printed maps
    were available.
    Using pattern recognition techniques of \href{https://urn.fi/URN:NBN:fi:aalto-201809034908}{\citet{masterthesis}},
    scans of the maps were converted into
    a digital format, which is compatible with maps of the categories A and B.
    The solar disk position as well as the rotational axis was already marked
    on these paper maps using a compass and a ruler.
  \end{enumerate}
Each line in a digital solar map file contains the time and coordinates of a
sample as well as the recorded intensity value. The coordinates are relative to
the expected location of the center of the Sun on the sky. The intensity is
measured on $\SI{37}{GHz}$ and contains linear and logarithmic output from the
amplifier as digital values. In this paper, we focus on the linear scale.
However, the logarithmic scale is more sensitive for very low intensities and
also avoids saturation during flare events, thus offering prospects for future
research.

We use the word \emph{calibration} for describing the combined effort of normalising and centering a map. The applicable
methods are described in Sect.~\ref{sect:disk}. The raw intensities of a scan have an arbitrary scale, as in
Fig.~\ref{typicalmap}{(a-c)}.  The signal levels must be normalised and also corrected for any minor offsets in the
sample coordinates in order to have a properly centered map.  From normalisation we obtain zero intensity at the
background and unit intensity for the QSL.  For the contour maps up to year 1987, we trust the original compass and
ruler markings and no additional calibration is considered.

  \subsection{Solar map centering and normalisation} \label{sect:disk}

Solar maps are generated by sweeping the beam of the radio telescope at the proximity of the solar disk while recording
signal intensity. In order to extract heliographic information from these maps, we first need to superimpose the visible
solar disk over these samples.  After correctly locating the disk, each antenna sample is accurately projected onto the
rotating heliographic surface.  The disk position is located by analysing the intensity distribution on the scans.  The
size of the visible disk is based on the distance between the Sun and the observer (Earth) at the time of observation.
We assume that the radius of the Sun is $\SI{717.6}{Mm}$.  This value is based on iterating the limb brightning model
using the $\SI{37}{GHz}$ MRO solar maps between 16th June, 2014 and 14th September, 2022, which are the three most
recently completed quarters of the solar cycle. We define the boundary of the heliographic sphere to be where its projection on the
solar map has intensity $0.5$ QSL.

Each solar map $S$ consists of $N_S$ samples $s_i$:
\eqnl{radio_sample}{ S &=& \left\{ s_i
\;:\; i = 1, 2, ..., N_S \right\}\,\text{ where}\nonumber\\
      s_i &=& \left( t_i,x_i, y_i, u_i, ... \right)\text{.}
}

Each sample $s_i$ contains the time of observation, $t_i$, and location on the
map, $(x_i, y_i)$, as well as the recorded intensity $u_i$. During the
calibration process described below, additional derived quantities are
attributed to each tuple $s_i$. These are denoted as $...$ on Eq.
\ref{radio_sample}. The calibration is iterative, with symbol $c$ used for each
step, starting with $c=0$. Each iterative step $c$ produces a set of
coordinates $(x_i^{(c)}, y_i^{(c)})$ and a calibrated intensity value
$v_i^{(c)}$, which equals zero at the background and unity at the average
interior disk brightness, which will be defined below. We denote the radial
distance from the center of the solar disk as: \eqnl{radio_radius}{ r_i^{(c)}
:= \sqrt{x_i^{(c) 2} + y_i^{(c) 2}} \text{.} }

Location is specified in equatorial coordinates relative to the expected
position, $(\RA_{\astrosun}, \Dec_{\astrosun})$, of the central point of Sun at
time $t_i$. We do not rely on the targeting accuracy of the antenna, and
instead determine the center of the visual disk from each map $S$, from which
we define the relative location of each $s_i$.

With right ascension cosine corrected, such that maps have unit aspect ratio,
we get the final coordinates (for large $c$) as:
\eqnl{relative_radec}{
  x_i^{(c)} = \frac{\RA(s_i)
      - \RA_{\astrosun}(t_i)}{\cos \left( \Dec_{\astrosun}(t_i) \right)}\text{,} \quad y_i^{(c)} = \Dec(s_i) - \Dec_{\astrosun}(t_i) \text{.}
}

The intensity is high at the solar disk, and it drops by orders of magnitude as
we transition to the background sky.  A number of factors affect the shape of
this transition:
\begin{enumerate}
\item Due to the shape and width of the beam used by the MRO $\SI{14}{m}$ disk,
near the limb convolution with the black sky reduces the observed solar disk
intensity.
\item The optical depth of the chromosphere varies with altitude due to
increasing temperature (\citeads{1978stat.book.....M}, \citeads{2017A&A...605A..78A}).  Nearer the limb, where the
observed line of sight makes a low angle through the corona, we would expect
increased intensity due to the higher temperatures compared to the center of
the disk.  We refer to this effect as limb brightening.
\item Observed intensity at $\SI{37}{GHz}$, as a time average over durations comparable to solar cycle length and butterfly diagram features (wings), can vary by heliographic latitude.
\end{enumerate}

The latitude dependence, which varies by solar cycle phase, is weak relative to beam convolution and limb brightening
effects. For map calibration purposes, we ignore these latitude dependency effects.  The importance of this
approximation is a subject for future study.  We have, however, implemented tools (see Appendix\,\ref{sect:limb}) for
taking into account the latitude dependency in order to measure certain chromosphere parameters. For this manuscript we
assume a radially symmetric antenna beam and thus the profile to correct for limb brightening is also radially
symmetric.

For each map, $S$, the raw signal intensity values, $u_i$, that come from the amplifier and analog-to-digital conversion, have an arbitrary scale.  These require a normalisation to extract two representative intensity
values, $\s{u}{background}$ and $\s{u}{disk}$. We have trialed three alternate methods for this initial step:
\begin{enumerate}
\item S-curve method, described orally by Juha Aatrokoski at MRO and used in MRO maps such as in \citetads{2018AN....339..204K}.
\item A variant of a convex hull method described in \citetads{masterthesis}.
\item Selecting peak values in intensity histogram, a method adopted from \citetads{2013PASJ...65S..17S}.
\end{enumerate}

Here we have applied method 1, which we shall describe in detail. The method for the initial step
is not crucial since the final result is derived by series of converging iterations that improve the centering and normalisation.
We do not rely only on the initial step. When there are bright or dim regions at the boundary, we discard them as outliers and limit the circle fitting to the remaining points.
While deriving the limb profile, as displayed in Fig.\ref{limb_brightening}(b), we can improve the estimate of a sample's radial distance based on its intensity, as illustrated in Fig.\ref{radio_sample}(c).

\begin{figure}
  \centering
  \includegraphics[width=8.5cm]{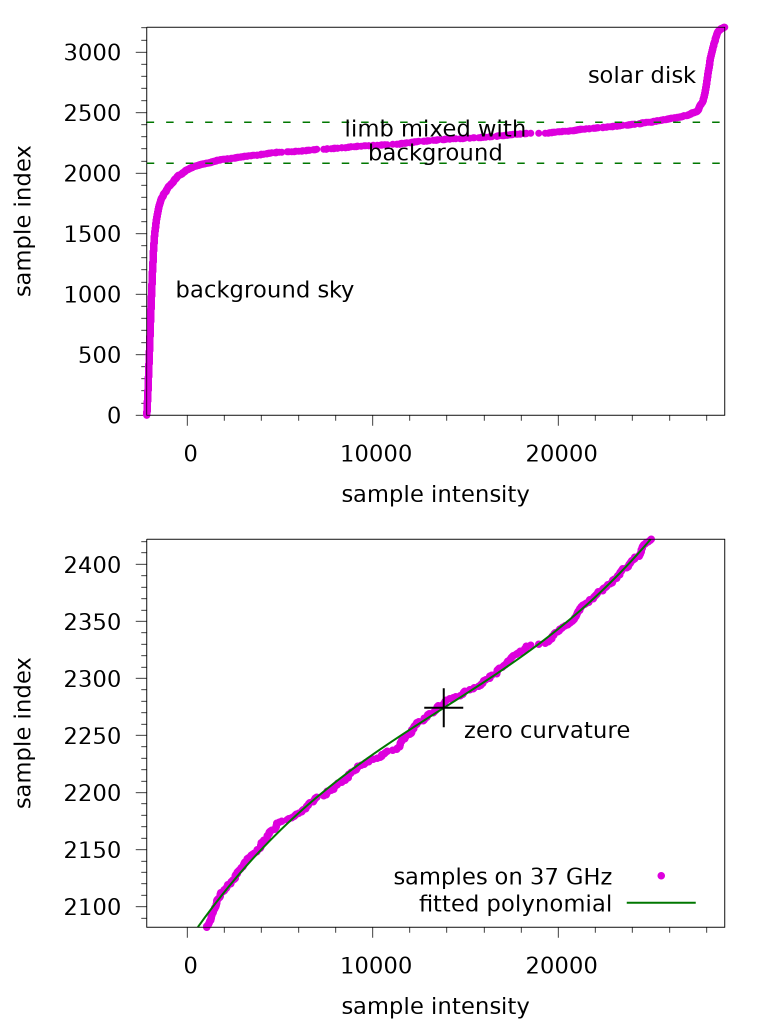}
  \begin{picture}(0,0)(0,0)
    \put(-245,308){\begin{large}{\sf\bf{(a)}}\end{large}}
    \put(-245,146){\begin{large}{\sf\bf{(b)}}\end{large}}
  \end{picture}
  \caption{
    Radio samples of a typical solar scan sorted according to intensity.  Panel
{(a)} displays all samples, within the limb transition between background and
disk located with the dashed lines.  A zoom in between the dashed line in {(b)}
is fitted with a polynomial, whose inflection (cross) is a pivot
point for signal level normalisation.}
  \label{S-curve_example}
  \end{figure}
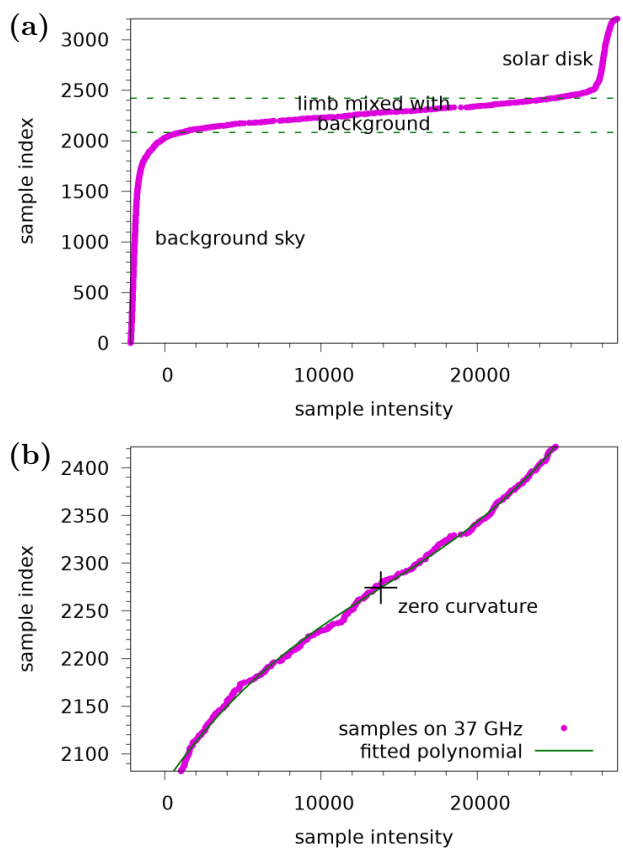

\begin{figure}
\centering
\includegraphics[trim=0.7cm 0.0cm 0.3cm 0.0cm,clip=true,width=8.5cm]{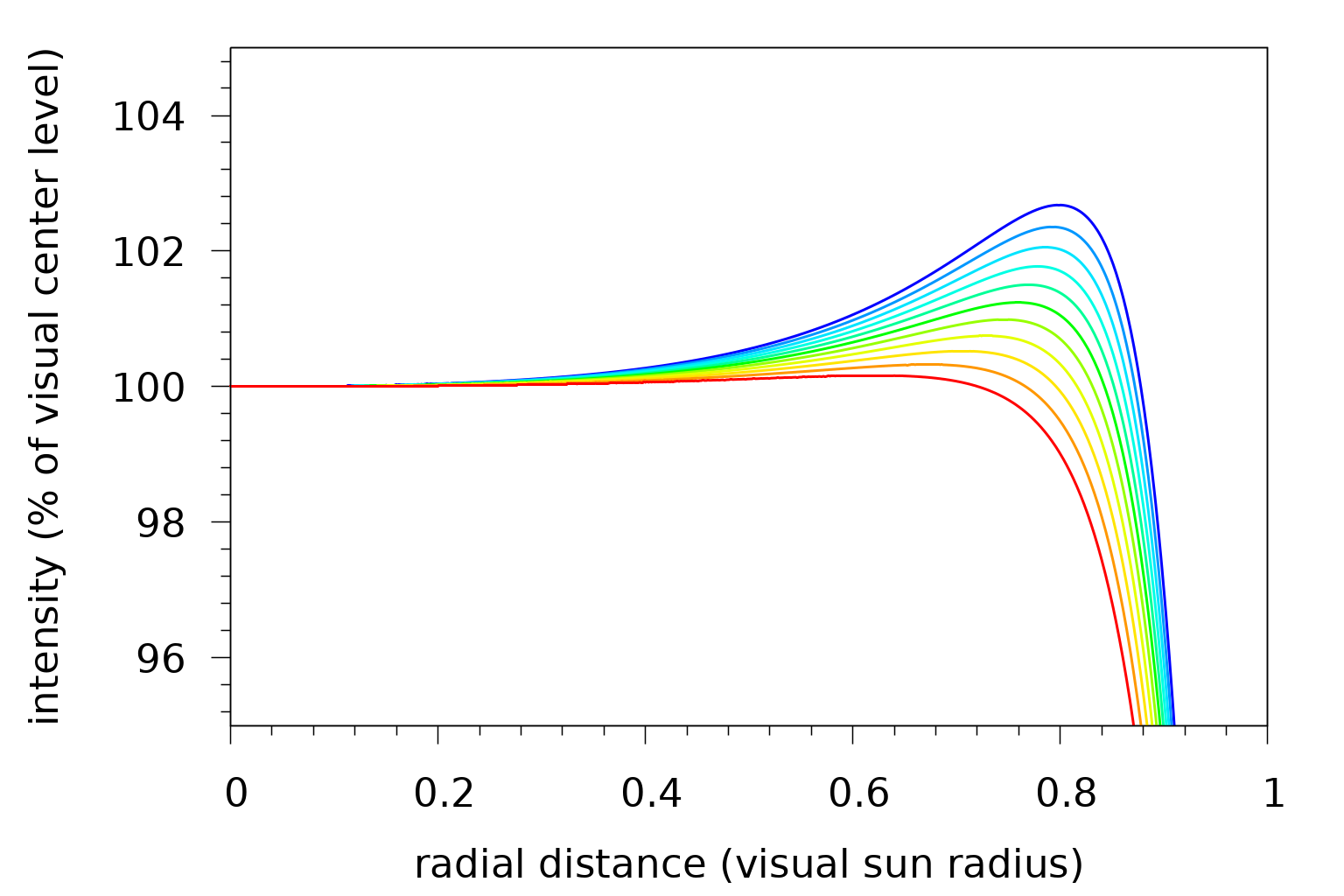}
\includegraphics[trim=0.5cm 0.0cm 0.5cm 0.8cm,clip=true,width=8.5cm]{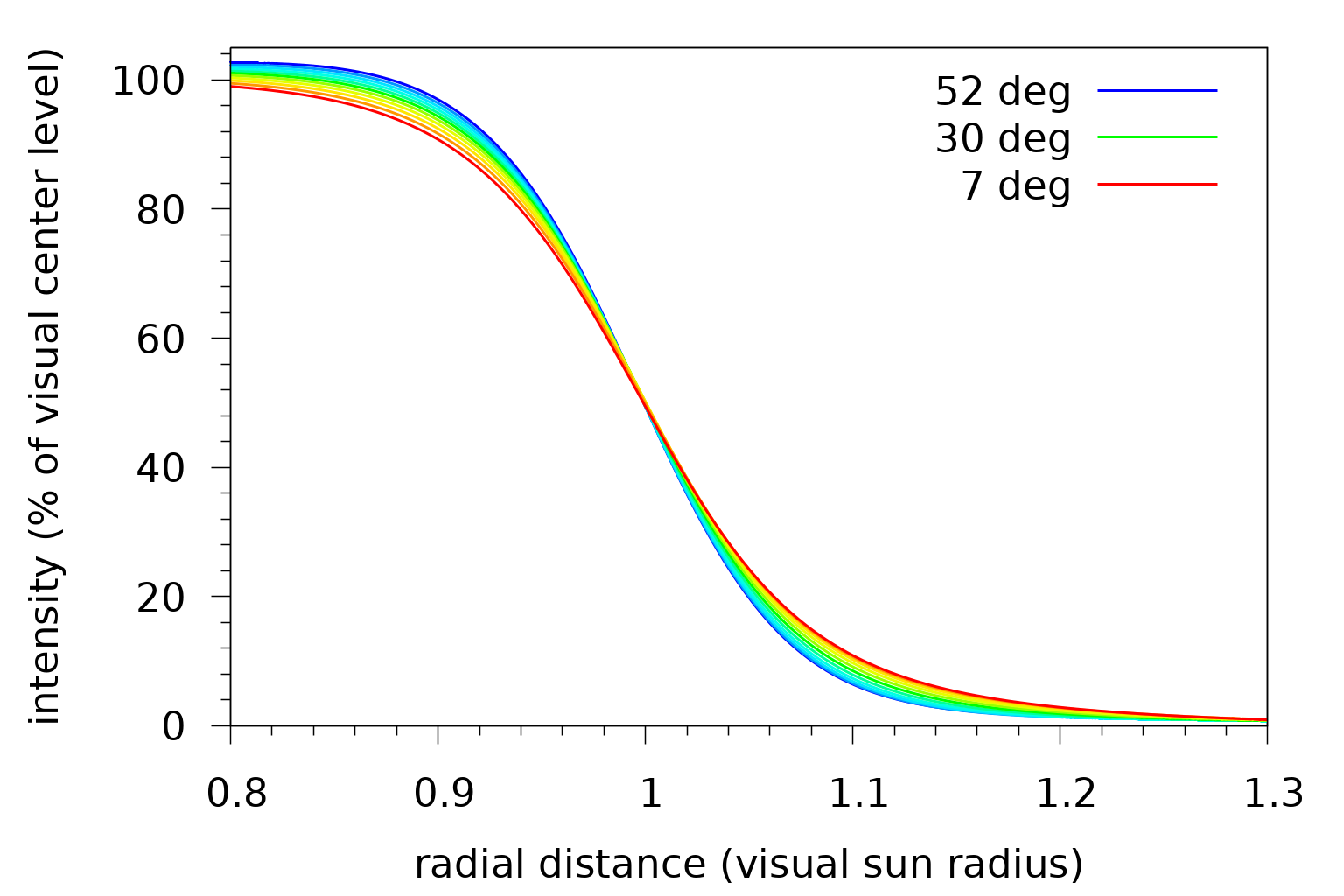}
  \begin{picture}(0,0)(0,0)
    \put(-200,295){\begin{large}{\sf\bf{(a)}}\end{large}}
    \put(-200,135){\begin{large}{\sf\bf{(b)}}\end{large}}
  \end{picture}
\caption{Average quiet Sun intensity as a function of radial distance from the
visual center, for different altitudes (elevations) of observation at MRO.
Fitted from data between 16th March, 2017 and 31st January, 2024. The 
profile varies over time, and models fitting each
quarter of a Solar cycle have been derived, which differ only slighly from the
displayed example. (a) Limb brightening is observed when line of sight is
almost tangential to the solar surface. (b) The beam size of $2.4^\prime$ mixes
the solar limb with the dark background, resulting in a convolution pattern. On
both plots, we observe more blurring when the altitude of observation is low,
due to atmospheric effects.}
\label{limb_brightening}
\end{figure}
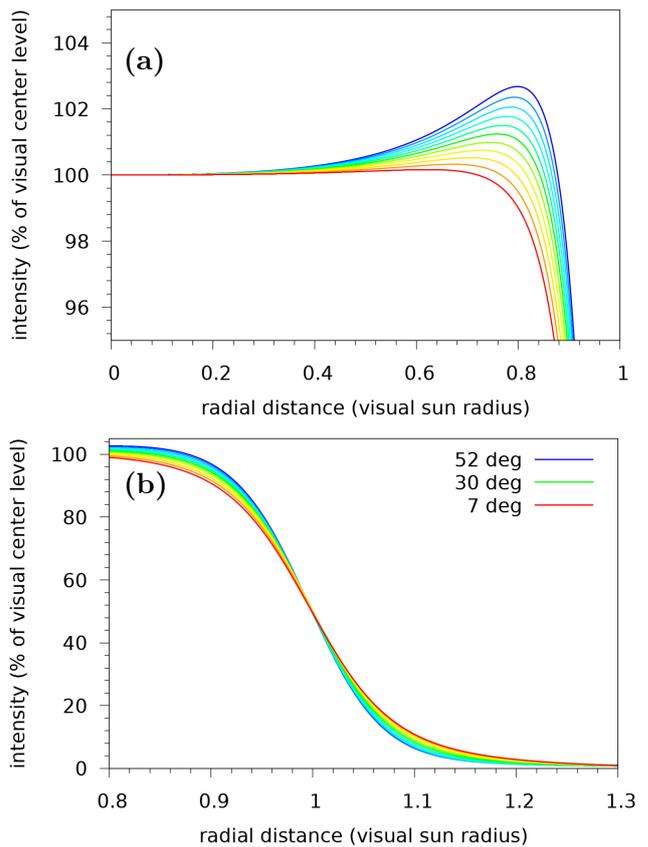

A through analysis and comparison of different calibration methods is a possible subject of further research. Following
method 1, the samples $s_i\in S$ are sorted by absolute intensity.  An example of a resulting intensity vs. sample
index is plotted in Fig.\,\ref{S-curve_example}{(a)}.  We observe two vertical regions in the plot, since the dark
background as well as the bright solar disk cover most of the samples. Between these regions, bounded by the dashed
lines, the curvature of the plot changes sign. Here we place the pivot point of the normalisation
(Fig.~\ref{S-curve_example}){(b)}.  The median intensity of samples with intensity higher than the pivot is taken as
representative of the inner disk. The lowest intensity tertile is taken as representative of the background. This fixes
the signal range from background to QSL as the starting point of further iterations, to identify the disk centre, radius
and intensity normalisation.

We then perform a linear transformation for the raw intensity values $u_i$ and
obtain the first normalised intensities $v_i^{(0)}$ as
\eqnl{map_normalisation}{
 v_i^{(0)} := \frac{u_i - \s{u}{background}}{\s{u}{disk} - \s{u}{background}} \; \text{with} \; s_i \in S \text{.}
}
The normalised intensities are then used for calculation of the initial center of the map, $(x^{(0)}, y^{(0)})$:
\eqnl{map_weighting}{
( x^{(0)}, y^{(0)} ) := \frac{\sum m_i * ( x_i, y_i )}{\sum m_i} \; \text{.}
}
The samples are weighted by $m_i$ as to prevent disk features from affecting the center of mass.
\eqnl{map_weighting2}{
m_i = \begin{cases}0 & \quad \text{when} \quad v_i \le 0.15 \\ \frac{v_i - 0.15}{0.7} & \quad \text{when} \quad 0.15 < v_i < 0.85 \\ 1 & \quad \text{when} \quad v_i \ge 0.85 \end{cases} \text{.}
}
This allows us to refine the equatorial coordinates of the samples as:
\eqnl{map_centering}{
(x_i^{(0)}, y_i^{(0)} ) := ( x_i, y_i ) - ( x^{(0)}, y^{(0)} )\; \text{.}
}
From this initial calibration round we iterate, by which we denote the superscript $^{(c)}$ with $c$ replacing 0 in
Eq.s~\ref{map_normalisation} and~\ref{map_centering}.  Each calibration round produces a new normalised intensity value
$v^{(c+1)}$ for each sample.  Using these new values $v^{(c+1)}$, the disk center is further optimised as $(x^{(c+1)},
y^{(c+1)})$. The iterations for producing superscripts $c > 0$ are based on fitting a circle to the boundary points,
outliers neglected.

Here we assume a fixed radius for the circle, based on known physical radius of the Sun as well as
known Earth-Sun distance. The physical radius of the Sun, for $\SI{37}{GHz}$, is defined by
convergence of iterating the limb profile for the whole MRO dataset in a so called grand iteration.
All maps are first calibrated, after which we fit one limb profile and temporal latitudinal
dependence for each subset of maps constituting three subsequent quarters of a solar cycle. Then we
solve the relative radius where the average intensity coincides $\SI{0.5}{QSL}$, and use that radius
to fine tune the calibration for a subsequent grand iteration. In that subsequent grand iteration,
the limb profile at $\SI{0.5}{QSL}$ will be closer to unit radii, and further converges in a few of
these grand iterations. In one grand iteration, any three adjacent quarters of a solar cycle produce
one limb profile and radius correction. For next grand iteration, we apply them to the middle
quarter and for possible boundary quarters that do not appear in the middle of any triplet of
adjacent quarters. These boundary quarters are the yet incomplete quarter 15th September, 2022 to
15th July, 2025 as well as the first quarter, 3rd March, 1989 to 30th August, 1991, to contain maps
in such a format that a limb model is applicable. Thus every quarter of a solar cycle has its own
radius and limbmodel. When producing a new limb profile and temporal latitudinal dependence as well
as a radius correction, preferably two more quarters, one on both sides, are used as to prevent
ringing artifacts in the middle quarter. The convergent radius for the most recent three complete
quarters (16th June, 2014 to 14th September, 2022) is $\SI{717.6}{Mm}$, representing the quarter
16th March, 2017 to 14th December to 16th March, 2019. For other quarters, the value fluctuates a
couple of megameters, and it fluctuates more for the cycles 22 and 23 which have sparser data. This
is a subject of further study.

  \subsection{Physical radius of the Sun}\label{sect:physical_radius}

  In this paper, we have assumed that the scale of the relative right ascension
and declination are correct and thus the radius of the disk is based on
geometry. Let $R_{\astrosun}$ be the physical radius of the Sun and $d(t)$ be
the distance between MRO on Earth and the physical center (core) of the Sun at
time $t$. Then:

\eqnl{visual_angle}{
r_{\astrosun}(t) = \arcsin \left( R_{\astrosun} / d(t) \right) \text{.}
}

The apparent radius of the Sun depends on wavelength and method of observation
as well as the exact definition. For radio wavelengths, a quadratic fit of the
apparent radius $\ave r$ is constructed in \citetads{2015ApJ...812...91R}.  Evaluating this
for $\SI{37}{GHz}$ ($\lambda = \SI{8.1}{mm}$ and the avarage distance of $\ave
d = \SI{1}{AU}$, we obtain $R_{\astrosun} = \SI{710.06}{Mm}$.
\citetads{2019SoPh..294..175S} examined the variation of the solar radius for
$\SI{37}{GHz}$ from MRO across cycles 22 to 24.  Instrumental changes resulted
in differences between upgrades in the measured radius between
979$^{\prime\prime}$ and 999$^{\prime\prime}$.  The values were based on
monthly averages of the radii, which were further averaged over the period of
instrumentation being unchanged. This will normalise for difference of the
apparent radius between perihelion and aphelion, which is ca.
$33^{\prime\prime}$.  The values in \citetads{2019SoPh..294..175S} thus translate to
physical diameters $\SI{710}{Mm}$ and $\SI{725}{Mm}$.

Here, we have analysed the solar radius using two different definitions that
apply to MRO solar maps. We use in this paper the experimental value
$R_{\astrosun} =\SI{717.6}{Mm}$, which provides $0.5$ QSL intensity at the disk
boundary when the altitude of the Sun is $25^{\circ}$ above the horizon during
observation. This definition is not actually very sensitive to the exact
altitude, as can be seen in Fig.\,\ref{limb_brightening}(b). However, we fix
the reference altitude in order to establish convergence. The limb profiles are
used for centering the solar maps in the next iteration, from which subsequent
and, eventually final, limb profiles are created. The processes of correcting
for limb brightening and convolution effects are detailed in Appendix~\ref{sect:limb}
and accounting for chromospheric effects along different lines of sight in
Appendix~\ref{sect:chrom}

\subsection{Heliographic coordinates from radio data}\label{sect:helio}

Our solar maps are not sensitive to small systematic pointing errors or
intermap variations in signal levels, since we calibrate each scan
individually.  Applying the iterative algorithm outlined in
Sect.~\ref{sect:disk} we obtain a calibrated position and intensity for each
sample, with levels scaled such that zero is for the background and unity for
the QSL:
\eqnl{calibration}{
v_i^{(c)} &=& \frac{u_i - \s{u}{background}^{(c)}}{\s{u}{QSL}^{(c)} - \s{u}{background}^{(c)}} \text{.}
}
We use an astronomical model, including the visible solar radius
$r_{\astrosun}$ and the observer's geolocation (Metsähovi Radio Observatory):
\eqnl{mro_geolocation}{
(\phi^{\earth}, \lambda^{\earth}) = \left( +60.22^{\circ}, +24.39^{\circ}
\right) \text{,}}
to determine the equatorial location of the centre of the solar disk as
observed, denoted by
$( \mathrm{RA}_{\astrosun}(t), \mathrm{dec}_{\astrosun}(t) )$.
Neglecting the motion of the Sun across the sky and the rotation of the Sun
within the duration of each scan $S$, for simplicity, we assume the middle time
\[\s{t}{mid}(S) = \frac{\s{t}{min}(S) + \s{t}{max}(S)}{2}.\] A simple
astronomical model is then used to project the cosine-corrected Sun-relative
equatorial coordinates $(x_i^{(c)},y_i^{(c)})$ into an idealised heliographic
surface set of Carrington coordinates
$(\phi_{\astrosun,i}, \lambda_{\astrosun,i})$.  This model assumes the Earth
in on a keplerian orbit around the Sun, and the rotation axis of Earth is
precessing at the period of ca. $\SI{26000}{a}$. Any inaccuracy in this
simple model will only arise a slight pointing error of the solar rotation
axis on the maps. Since the antenna beam radius is large ($2.4^\prime$) this
simple astronomical model as adequate.

\section{Results}\label{sect:results}

\begin{figure*}
\centering
\includegraphics[trim= 0.00cm 4.4cm  0.0cm 0.2cm,clip=true,width=17cm]{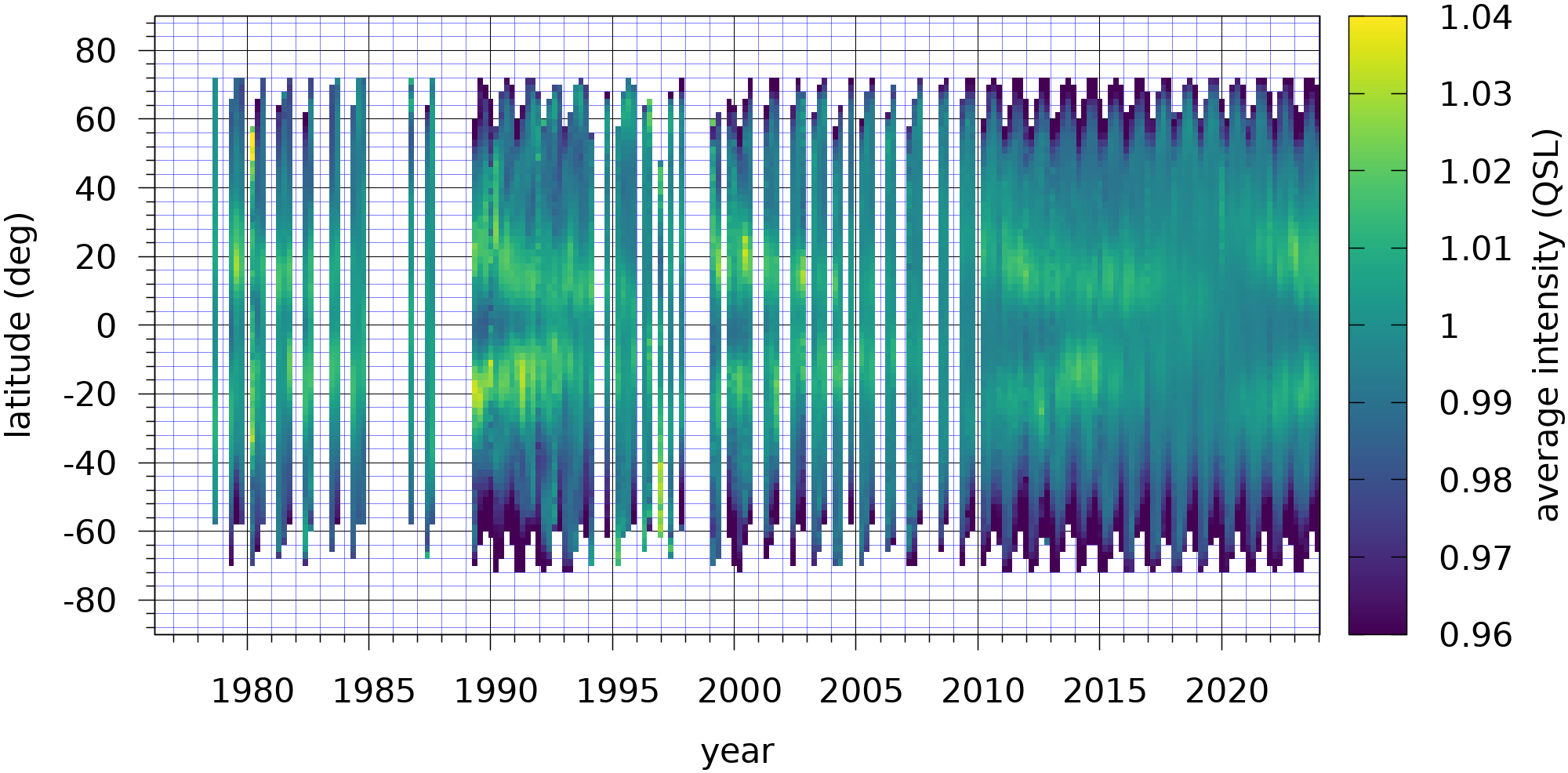}\\
\includegraphics[trim= 0.00cm 4.4cm  0.0cm 0.2cm,clip=true,width=17cm]{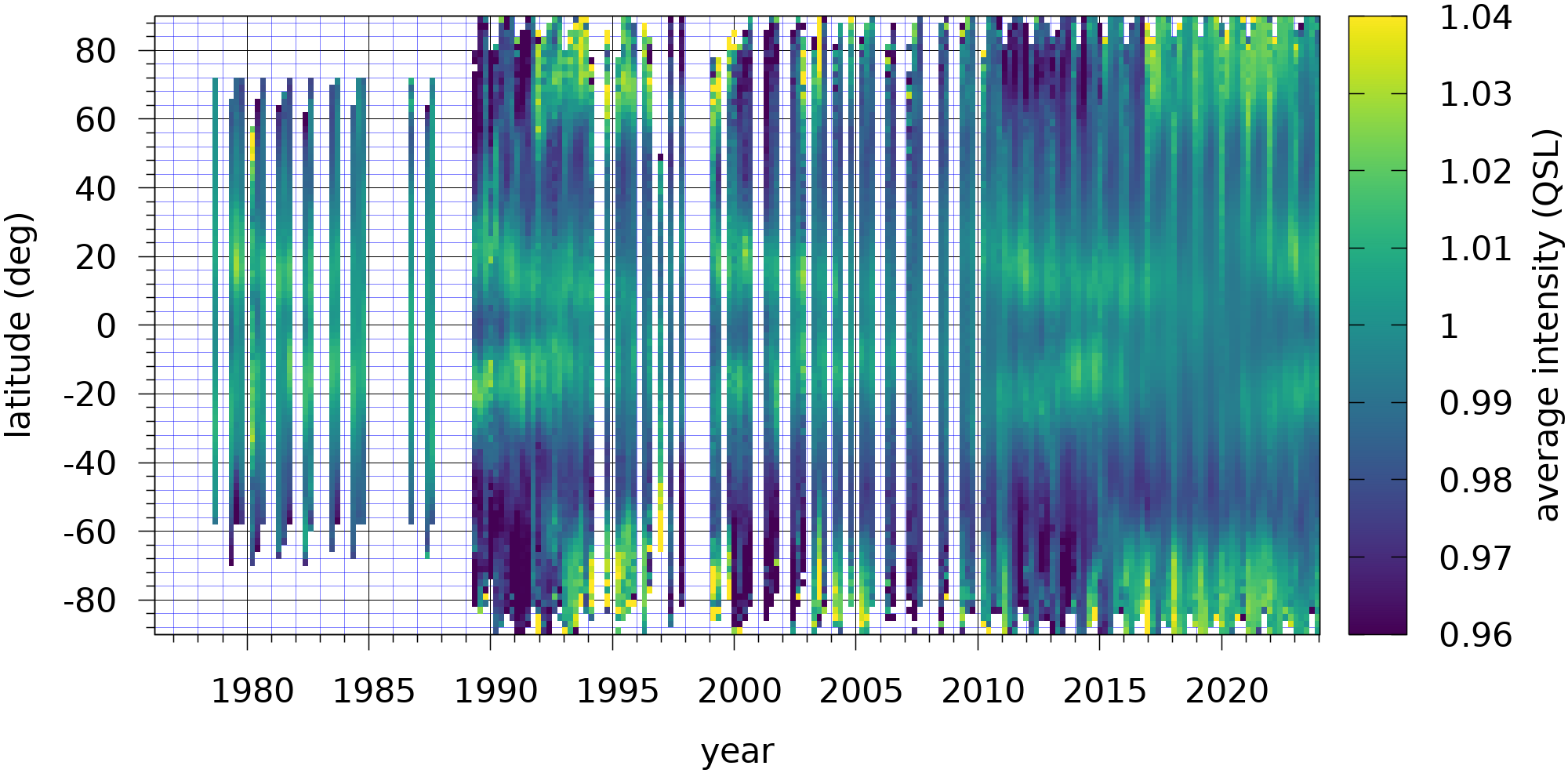}
\includegraphics[trim=-1.54cm 0.0cm -3.4cm 0.0cm,clip=true,width=17cm]{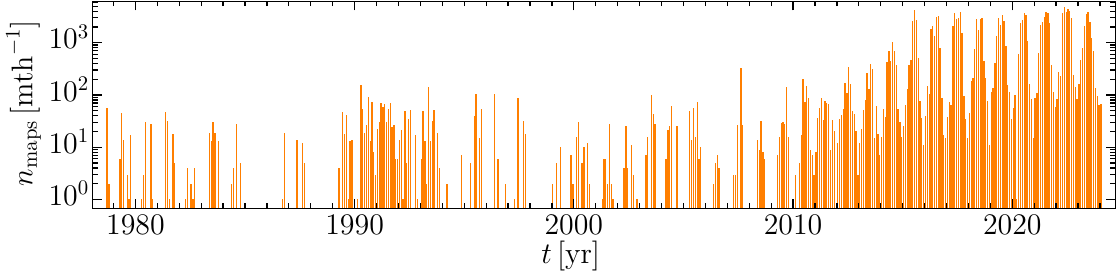}
\begin{picture}(0,0)(0,0)
  \put(-488,490){\color{black}{\bf{(a)}}}
  \put(-488,285){\color{black}{\bf{(b)}}}
  \put(-488, 80){\color{black}{\bf{(c)}}}
\end{picture}
\caption{Solar intensity at all heliographic latitudes on $\si{37}{GHz}$ over
cycles 21 to 24. Averaged over samples for which (a) $r \le 0.9r_{\astrosun}$
to avoid beam mixing with the background, and (b) $r \le 1.0r_{\astrosun}$, such
that each sample intensity is corrected for beam mixing and limb brightening.
The latter increases the available latitude range, revealing the polar
magnetic cycle.
(c) Count of Metsähovi solar maps used for analysis after being accepted by automated quality control.
\label{maps_per_month}
\label{butterfly_clear_corr}
\label{butterfly_clear_raw}}
\end{figure*}
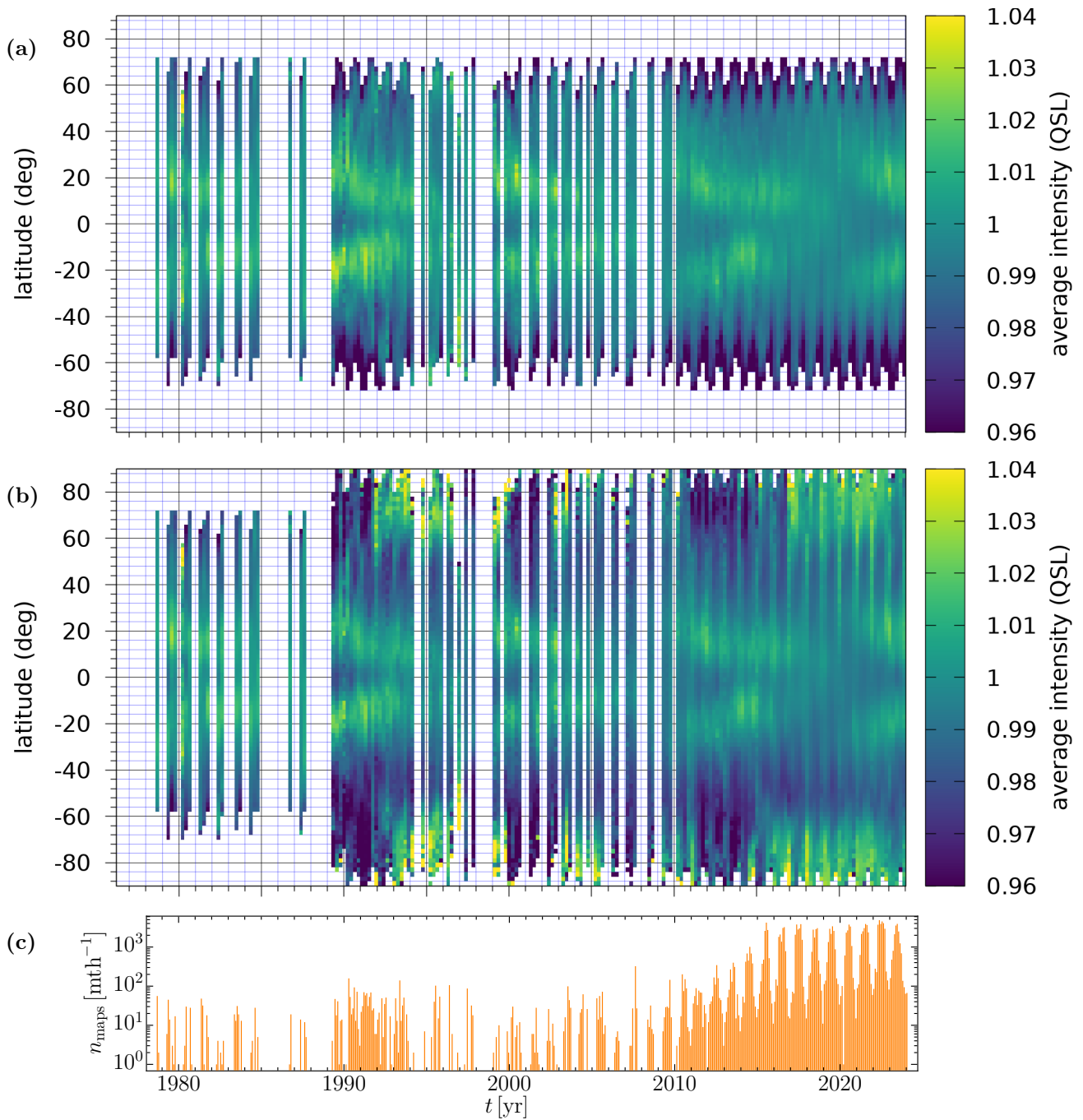

After disk fitting, each radio sample can be projected onto the Carrington
coordinates of the rotating heliographic surface.  Every sample now has time,
intensity relative to QSL and heliographic latitude.  We combine these to
construct a time-latitude butterfly diagram.  From the series of radio maps
using processing described in Sect.~\ref{sect:methods}, but without applying
the limb correction model (Appendix~\ref{sect:limb}), we construct the radio
butterfly diagram displayed in Fig.~\ref{butterfly_clear_raw}{(a)}.

Fig.~\ref{butterfly_clear_raw}{(a)} shows the map radio samples averaged over a
grid of time and heliographic latitude. In the grid, there are $90$ latitude
bins, ranging $\pm 90^{\circ}$ of heliographic latitude, and $220$ time bins
ranging from March, 1974 to January, 2024. The sample intensities are normalised
as described in \ref{sect:disk} so that for each sample, the intensity is
relative to the statistical Quiet Sun Level of the corresponding map to which
the sample belongs.

We present two versions of the butterfly diagram. Fig.~\ref{butterfly_clear_raw}(a) has no limb
correction, omitting samples that have $r > 0.9r_{\astrosun}$ and thus losing information from the
polar regions. Fig.~\ref{butterfly_clear_corr}(b) applies the limb correction model identified in
Fig.~\ref{limb_brightening}, and Appendix~\ref{sect:limb}. Comparing the two, it is clear that
without the limb correction model, even for samples at $0.8r_{\astrosun} < r_i < 0.9r_{\astrosun}$
the intensities are lower than physical values due to beam mixing with the background.
Data are binned according to heliographic latitude and time.  The value for each bin is an average of the samples belonging to the bin.

Our time series now extends the previously published data to include the
observations digitally recovered from the historical maps covering cycle 21.
The imbalance in the statistics available across the time span is evident in
more sparse observations during earlier epochs and the seasonal variation in
quantity and quality of the specimens as presented in
Fig.\,\ref{maps_per_month}(c).

\section{Summary and discussion}\label{sect:discussion}

Without the limb correction our butterfly diagram
Fig.\,\ref{butterfly_clear_raw}(a) during 1993 to 2013 recovers the same
activity pattern for latitudes $<|50^\circ|$ as \citetads[][see their
figure~2]{2013PASJ...65S..17S} at $\SI{17}{GHz}$, see also
(\citeads{2017JKAS...50..125K}, \citeads{2016ApJ...823L..15G}) and also
for $\SI{5.7}{GHz}$ \citepads{2016ASPC..504...71P}.  However, only
with the limb correction does our Fig.\,\ref{butterfly_clear_corr}(b) also
reproduce the high intensities approaching polar latitudes.  For both
observatories the activity at the Southern pole is more intense and prolonged,
and while it appears to cease in 2012 in the earlier work, from our results we
see that this is only a temporary lull and that this activity has a much longer
duration. The results of \citetads{2013PASJ...65S..17S} only date from 1993,
during which the poles show the onset of only enhanced activity. Our results
including 1989 indicate this was preceeded by a a period on consistent reduced
polar magnetic activity, which may indicate that the period of the polar
magnetic cycle exceeds the 11 year solar cycle.  Extending the observations at
$\SI{17}{GHz}$ to 2018 \citetads{2018JASTP.176...26G} the activity at the
Northern pole reduces after 2013, consistent with our presentation
\citepads[see also][]{2019SoPh..294...30F}. There is a risk that the signal
near the limb may be corrupted by noise from limb brightening and convolution
within the beam width of limb and background intensity, but the long-term
persistence of weak followed by strong intensities at the poles and consistency
with similar microwave observations supports the conclusion that these effects
are physical rather than instrumental.

\citetads{2018FrASS...5...38S} explore the properties of the extended solar
cycle (ESC) to around 17 years preceding the 11 year cycle by about 8 years,
which appears to be a residue of earlier cycles evident at higher heliographic
latitude and higher solar atmospheric phenomena. The extended cycle is
identified in the Fe XIV green line emissions in the solar corona by
\citetads{1997SoPh..170..411A} and further analysis using SDO/HMI data
\citepads{2014ApJ...792...12M} supports the hypothesis that the magnetic
processes leading to the emergence of each new sunspot cycle have observational
signatures apparent a few years in advance. The extended period of high
intensity near the poles apparent in Fig.\,\ref{butterfly_clear_corr}(b) may
be relevant to understanding the ESC phenomenon.

North-South asymmetry with a lag as much as two years is identified by
\citetads{1955MNRAS.115..398N} and is evident in asymmetry of our butterfly
wings, particularly for cycles 23 and 24, and in the polar activity after 2010.
\citetads{2023A&A...674A.182M} uses $\SI{2.8}{GHz}$ data to identify a
North-South asymmetry in the overall flux across multiple odd cycles, as
evidence of the presence of a relic field, coexisting with the dynamo driven
magnetic field. As discussed above, there are significant inbuilt biases in the
measurements with respect to North and South poles, so further work would need
to be applied to our data, before we could determine whether or not our
observations support such conclusions.

In summary, we have presented one of the longest continual measurements at
$\SI{37}{GHz}$ of solar chromespheric activity available, having applied
corrections for limb brightening and convolution effects and atmospheric and
orbital biases, providing an accurate historical catalog of these observations.
Our processing of the raw data has permitted us to include measurements near
the solar poles, which indicate a prolonged period of magnetic activity which
may exceed the 11 year solar cycle. The individual maps contain regions of
increased intensity that persist over days and weeks and indicated enhanced
magntic activity. Future work shall present automatic feature recognition and
activity tracking tools and results in order to relate these to magnetic
activity in the chromosphere and elsewhere.

\begin{acknowledgements}
 This publication makes use of data obtained at the Metsähovi Radio Observatory, operated by the Aalto University.
 F.A.G. and M.J.K. acknowledge support from the Academy of Finland
 ReSoLVE Centre of Excellence (grant 307411) and the ERC
 under the EU's Horizon 2020 research and innovation
 programme (Project UniSDyn, grant 818665).
 S.K.K. acknowledges support from Niilo Helander Foundation (awarded 2018) as well as Jenny and Antti Wihuri Foundation (awarded 2019).
\end{acknowledgements}

\bibliographystyle{aa}
\bibliography{main_P1}

\begin{thebibliography}{32}
\expandafter\ifx\csname natexlab\endcsname\relax\def\natexlab#1{#1}\fi

\bibitem[{{Alissandrakis} {et~al.}(2017){Alissandrakis}, {Patsourakos},
  {Nindos}, \& {Bastian}}]{2017A&A...605A..78A}
{Alissandrakis}, C.~E., {Patsourakos}, S., {Nindos}, A., \& {Bastian}, T.~S.
  2017, \aap, 605, A78

\bibitem[{{Altrock}(1997)}]{1997SoPh..170..411A}
{Altrock}, R.~C. 1997, \solphys, 170, 411

\bibitem[{{Babcock}(1961)}]{1961ApJ...133..572B}
{Babcock}, H.~W. 1961, \apj, 133, 572

\bibitem[{{Fujiki} {et~al.}(2019){Fujiki}, {Shibasaki}, {Yashiro}, {Tokumaru},
  {Iwai}, \& {Masuda}}]{2019SoPh..294...30F}
{Fujiki}, K., {Shibasaki}, K., {Yashiro}, S., {et~al.} 2019, \solphys, 294, 30

\bibitem[{{Gaizauskas} {et~al.}(1983){Gaizauskas}, {Harvey}, {Harvey}, \&
  {Zwaan}}]{1983ApJ...265.1056G}
{Gaizauskas}, V., {Harvey}, K.~L., {Harvey}, J.~W., \& {Zwaan}, C. 1983, \apj,
  265, 1056

\bibitem[{{Gopalswamy} {et~al.}(2018){Gopalswamy}, {M{\"a}kel{\"a}}, {Yashiro},
  \& {Akiyama}}]{2018JASTP.176...26G}
{Gopalswamy}, N., {M{\"a}kel{\"a}}, P., {Yashiro}, S., \& {Akiyama}, S. 2018,
  Journal of Atmospheric and Solar-Terrestrial Physics, 176, 26

\bibitem[{{Gopalswamy} {et~al.}(2016){Gopalswamy}, {Yashiro}, \&
  {Akiyama}}]{2016ApJ...823L..15G}
{Gopalswamy}, N., {Yashiro}, S., \& {Akiyama}, S. 2016, \apjl, 823, L15

\bibitem[{{Hale} {et~al.}(1919){Hale}, {Ellerman}, {Nicholson}, \&
  {Joy}}]{1919ApJ....49..153H}
{Hale}, G.~E., {Ellerman}, F., {Nicholson}, S.~B., \& {Joy}, A.~H. 1919, \apj,
  49, 153

\bibitem[{{Kallunki} {et~al.}(2012){Kallunki}, {Lavonen}, {J{\"a}rvel{\"a}}, \&
  {Uunila}}]{2012BaltA..21..255K}
{Kallunki}, J., {Lavonen}, N., {J{\"a}rvel{\"a}}, E., \& {Uunila}, M. 2012,
  Baltic Astronomy, 21, 255

\bibitem[{{Kallunki} \& {Riehokainen}(2012)}]{2012AN....333...20K}
{Kallunki}, J. \& {Riehokainen}, A. 2012, Astronomische Nachrichten, 333, 20

\bibitem[{{Kallunki} {et~al.}(2020){Kallunki}, {Tornikoski}, \&
  {Bj{\"o}rklund}}]{2020SoPh..295..105K}
{Kallunki}, J., {Tornikoski}, M., \& {Bj{\"o}rklund}, I. 2020, \solphys, 295,
  105

\bibitem[{{Kallunki} {et~al.}(2018){Kallunki}, {Tornikoski}, {Tammi},
  {Kinnunen}, {Korhonen}, {Kes{\"a}l{\"a}inen}, \&
  {Arkko}}]{2018AN....339..204K}
{Kallunki}, J., {Tornikoski}, M., {Tammi}, J., {et~al.} 2018, Astronomische
  Nachrichten, 339, 204

\bibitem[{{Kim} {et~al.}(2017){Kim}, {Park}, \& {Kim}}]{2017JKAS...50..125K}
{Kim}, S., {Park}, J.-Y., \& {Kim}, Y.-H. 2017, Journal of Korean Astronomical
  Society, 50, 125

\bibitem[{Kivistö(2018)}]{masterthesis}
Kivistö, S. 2018, Master's thesis, Aalto University

\bibitem[{{Krause} \& {Raedler}(1980)}]{1980mfmd.book.....K}
{Krause}, F. \& {Raedler}, K.~H. 1980, {Mean-field magnetohydrodynamics and
  dynamo theory}

\bibitem[{{Leussu} {et~al.}(2017){Leussu}, {Usoskin}, {Senthamizh Pavai},
  {Diercke}, {Arlt}, {Denker}, \& {Mursula}}]{2017A&A...599A.131L}
{Leussu}, R., {Usoskin}, I.~G., {Senthamizh Pavai}, V., {et~al.} 2017, \aap,
  599, A131

\bibitem[{{Loukitcheva} {et~al.}(2019){Loukitcheva}, {White}, \&
  {Solanki}}]{2019ApJ...877L..26L}
{Loukitcheva}, M.~A., {White}, S.~M., \& {Solanki}, S.~K. 2019, \apjl, 877, L26

\bibitem[{{McIntosh} {et~al.}(2014){McIntosh}, {Wang}, {Leamon}, {Davey},
  {Howe}, {Krista}, {Malanushenko}, {Markel}, {Cirtain}, {Gurman}, {Pesnell},
  \& {Thompson}}]{2014ApJ...792...12M}
{McIntosh}, S.~W., {Wang}, X., {Leamon}, R.~J., {et~al.} 2014, \apj, 792, 12

\bibitem[{{Mets\"ahovi Radio Observatory}(2019)}]{mrosolardata}
{Mets\"ahovi Radio Observatory}. 2019, Metsähovi Radio Observatory public
  solar database,
  \url{http://urn.fi/urn:nbn:fi:att:f371cb6d-f84c-4d76-99e4-c39c639fd0de}

\bibitem[{{Mihalas}(1978)}]{1978stat.book.....M}
{Mihalas}, D. 1978, {Stellar atmospheres}

\bibitem[{{Mursula}(2023)}]{2023A&A...674A.182M}
{Mursula}, K. 2023, \aap, 674, A182

\bibitem[{{Newton} \& {Milsom}(1955)}]{1955MNRAS.115..398N}
{Newton}, H.~W. \& {Milsom}, A.~S. 1955, \mnras, 115, 398

\bibitem[{{Pelt} {et~al.}(2006){Pelt}, {Brooke}, {Korpi}, \&
  {Tuominen}}]{2006A&A...460..875P}
{Pelt}, J., {Brooke}, J.~M., {Korpi}, M.~J., \& {Tuominen}, I. 2006, \aap, 460,
  875

\bibitem[{{Pevtsov}(2016)}]{2016ASPC..504...71P}
{Pevtsov}, A.~A. 2016, in Astronomical Society of the Pacific Conference
  Series, Vol. 504, Coimbra Solar Physics Meeting: Ground-based Solar
  Observations in the Space Instrumentation Era, ed. I.~{Dorotovic}, C.~E.
  {Fischer}, \& M.~{Temmer}, 71

\bibitem[{{Rozelot} {et~al.}(2015){Rozelot}, {Kosovichev}, \&
  {Kilcik}}]{2015ApJ...812...91R}
{Rozelot}, J.~P., {Kosovichev}, A., \& {Kilcik}, A. 2015, \apj, 812, 91

\bibitem[{{Selhorst} {et~al.}(2014){Selhorst}, {Costa}, {Gim{\'e}nez de
  Castro}, {Valio}, {Pacini}, \& {Shibasaki}}]{2014ApJ...790..134S}
{Selhorst}, C.~L., {Costa}, J.~E.~R., {Gim{\'e}nez de Castro}, C.~G., {et~al.}
  2014, \apj, 790, 134

\bibitem[{{Selhorst} {et~al.}(2019){Selhorst}, {Kallunki}, {Gim{\'e}nez de
  Castro}, {Valio}, \& {Costa}}]{2019SoPh..294..175S}
{Selhorst}, C.~L., {Kallunki}, J., {Gim{\'e}nez de Castro}, C.~G., {Valio}, A.,
  \& {Costa}, J. E.~R. 2019, \solphys, 294, 175

\bibitem[{{Shibasaki}(2013)}]{2013PASJ...65S..17S}
{Shibasaki}, K. 2013, \pasj, 65, S17

\bibitem[{{Srivastava} {et~al.}(2018){Srivastava}, {McIntosh}, {Arge},
  {Banerjee}, {Dikpati}, {Dwivedi}, {Guhathakurta}, {Karak}, {Leamon},
  {Matthew}, {Munoz-Jaramillo}, {Nandy}, {Norton}, {Upton}, {Chatterjee},
  {Mazumder}, {Rao}, \& {Yadav}}]{2018FrASS...5...38S}
{Srivastava}, A.~K., {McIntosh}, S.~W., {Arge}, N., {et~al.} 2018, Frontiers in
  Astronomy and Space Sciences, 5, 38

\bibitem[{{Steenbeck} {et~al.}(1966){Steenbeck}, {Krause}, \&
  {R{\"a}dler}}]{1966ZNatA..21..369S}
{Steenbeck}, M., {Krause}, F., \& {R{\"a}dler}, K.-H. 1966, Zeitschrift
  Naturforschung Teil A, 21, 369

\bibitem[{{Valtaoja} {et~al.}(1987){Valtaoja}, {Sillanpaa}, \&
  {Valtaoja}}]{1987A&A...184...57V}
{Valtaoja}, L., {Sillanpaa}, A., \& {Valtaoja}, E. 1987, \aap, 184, 57

\bibitem[{{Vecchio} {et~al.}(2012){Vecchio}, {Laurenza}, {Meduri}, {Carbone},
  \& {Storini}}]{2012ApJ...749...27V}
{Vecchio}, A., {Laurenza}, M., {Meduri}, D., {Carbone}, V., \& {Storini}, M.
  2012, \apj, 749, 27

\end{thebibliography}

\begin{appendix}
\section{Correction for limb brightning and antenna convolution}\label{sect:limb}

Beam width of the telescope defines the sharpness of the transition from ca.
$1$ QSL to background level (zero) seen in Fig.~\ref{limb_brightening}.
Diameter of the dish would suggest a beam diameter $2.4^\prime$, but this is
further distorted by atmospheric scattering on Earth. Thus the beam width and
shape is also a function of the altitude angle above horizon as well as local
weather.

As discussed in the beginning of Sect.\,\ref{sect:disk}, signal intensity near
the limb region is affected by beam convolution of the disk and background as
well as increased chromospheric thermal emission along the line of sight. The
small angle of separation between the tangent surface and the chromosphere at
the limb causes the observer to probe higher altitudes of the chromosphere,
where the temperature of the emitting plasma is higher.  This produces the limb
brightning, which is seen on Fig.\,\ref{limb_brightening}\,(a) as intensities
above $1$ QSL.

We have developed a model to compensate for the effects of limb brightening
and antenna convolution.  The model also allows for solar cycle variation
and terrestrial atmospheric scattering.  Signals observed from the Sun at
very low altitude travel a longer path through the air than from higher
altitudes, resulting in greater scattering.

The model is iterative, so that we have a set of parameters $M^{(c)}$ for
each step $c$ of the iteration. The parameter set $M^{(c)}$ contains
coefficients of two separate polynomials, $P^{(c)}$ and $Q^{(c)}$:
\eqnl{model_def}{
M^{(c)} = \left( P^{(c)}, Q^{(c)} \right) \text{.}
}
Both polynomials have two-dimensional domain set. The product of polynomials
$P^{(c)}$ and $Q^{(c)}$ is an estimate for the observed intensity based on
four variables:
\begin{itemize}
  \item $t$, time, phase of solar cycle.
  \item $l$, heliographic latitude of the intersection of observation line of sight and the heliographic surface.
  \item $r$, distance from the center of disk, relative to the apparent radius of the solar disk.
  \item $a$, altitude of observation, angle between observation line of sight and the horizon at MRO.
\end{itemize}

The model must capture the detail at the solar limb, which is, in practice,
when $r \in [0.8 r_{\astrosun}, 1.2 r_{\astrosun}]$. This detail originates
from the convolution by the antenna beam. We map $r$ into adjusted variable
$\rho$ using a tuning factor $\kappa$:

\eqnl{model_kappa}{
\rho(r) = \arctan \left( \kappa \left( \frac{r}{r_{\astrosun}} - 1 \right) \right) \; \text{with} \; \kappa = 8.8 \text{.}
}
The tuning factor $\kappa$ is set by manually searching for a value which provides the best results.

The estimated intensity, $v$, is then calculated as:
\eqnl{model_func}{
v = M^{(c)}(t,l,r,a) = P^{(c)}(t,l) \cdot Q^{(c)}(\rho(r),a) \text{.}
}
The model is optimised using the least squares method in order to minimize the error between actual observed intensities
and the estimates of the model. As a target set we use the antenna samples of solar cycle 24.

In Fig.\,\ref{limb_brightening}\,(a) radial profiles for limb brightening are
plotted for a range of heliographic latitudes.
Fig.\,\ref{limb_brightening}\,(b) for the same range shows the convolution
profiles.

The accuracy of the calibration and centering of the disk is improved by
applying the limb correction to the samples before each iteration round in Sect.\,\ref{sect:disk}.
It is also of value in determining the radius of the Sun, as described in
Sect.\,\ref{sect:physical_radius}.

We start with a crude model for the solar disk in order to give each antenna
sample the variables $l$ and $r$ needed in Eq.\,\ref{model_func}. We then
optimize for the first step of the iteration to get $M^{(1)}$.  Thus, the
estimated intensity and the distance from the center of disk are
interdependent. When we have the intensity, we can estimate radial distance as:
\eqnl{model_dist}{
Q^{(c)}(\rho(r),a) = \frac{v}{P^{(c)}(t,l)} \text{.}
}

Given Eq.\,\ref{model_dist}, we can extract $r$ by first substituting $a$
into the polynomial $Q^{(c)}$ to get one-dimensional polynomial
$Q_{a}^{(c)}(\rho)$. We then use its inverse function to evaluate $\rho$ and
$r$:
\eqnl{model_inv}{
r = \left( \rho^{-1} \cdot Q_{a}^{(c) -1} \right) \left( \frac{v}{P^{(c)}(t,l)} \right) \text{.}
}

The solar map centering is an optimisation problem, which seeks the circle to
best fit the set of limb samples. Using Eq. \ref{model_inv}, we can improve the
fitting using the information of how far from the center each limb sample
actually might be. We initially have a circle, centered at $C$, which fits to
the limb samples. We will now translate each sample radially, with respect to
$C$, to a distance set by Eq.\,\ref{model_inv}.

This is illustrated in Fig.\,\ref{typicalmap}(c), where the blue dots are true
observed limb samples. The corresponding green dots are translated to a
distance given by Eq.\,\ref{model_inv}. The set of green points provide a less
dispersed target set for fitting a circle. This causes a slight adjustment in
the center $C$, so we will iterate the fitting for a suitable number of rounds.

The accuracy of the calibration and centering of the disk is improved by
applying the limb correction to the samples before each iteration round in
Sect.\,\ref{sect:disk}.  It is also of value in determining the radius of the
Sun, as described in Sect.\,\ref{sect:physical_radius}.

The limb correction for map centering is illustrated in
Fig.\,\ref{typicalmap}\,(c). While most observed positions (antenna samples)
are denoted with white dots, the subset at the limb region is shown blue.  Each
of these blue dots has an arrow pointing towards a corresponding green dot.
Arrows indicate translations associated with the limb correction. The
translation is radial; if the observed limb sample is more than $0.5\,
\textrm{QSL}$, we translate it towards the center, and if it is less than
$0.5\, \textrm{QSL}$, we translate it outwards. The translations produce the
set of green points. The green points are the target set for fitting a circle,
which becomes the disk circumference and is marked with magenta color. We
observe that the green dots appear closer to the final boundary circle,
compared to the blue dots. Since there is less deviation, we expect a more
accurate result for circle fitting when we use the green dots as a target set.

We note that the interpolation (green) is slightly offset from the
circumference (magenta). The limb correction model compensates variations
associated to altitude of observation. The altitude $a$ relates to
line-of-sight optical thickness $x(a)$ of Earth's atmosphere as:
\eqnl{atmosphere}{
x(a) = \frac{x(90^{\circ})}{\sin a} \text{.}
}
The optical thickness of the air mass along line of sight is lightest when the
Sun is seen high. It is reasonable to assume that the amount of air mass is
significant to the shape of the antenna beam. However, there are other effects
as well, which are not taken into account in this simple model. For example
rain, snow, and turbulence will scatter the beam. The statistical fact of
having winter observations only at low altitudes is taken into account by
normalising the data for altitude.

In coming research, we will test the limb correction on individual maps along
with a more sophisticated observation model. In Fig.\,\ref{typicalmap}\,(c), we
see the antenna path wiggling along the scanlines. Once the weather effects are
compensated, we will verify the antenna position by analyzing whether the
reported wiggling is consistent with the expected intensity variations due to
the beam pointing slightly towards or away to the disc center at a given
sample.

The limb model is based on fitting a polynomial function on the set of all samples from all accepted
solar maps. Since there are various biases in amount of data, we apply several equalization steps
for the weighting of the samples. This is to avoid ringing artifacts in the limb correction model.

\begin{enumerate}[(i)]

\item The number of samples per map varies. Prior to June 2015 samples
were more densely recorded, also taking much longer to scan, as
illustrated by Fig.~\ref{oldmap}\emph{(a)}. Subsequently the samples are
recorded more rapidly with fewer samples required per map as shown in
Fig.~\ref{oldmap}\emph{(b)}. The number of samples also varies annually,
since the Sun looks larger when Earth is at perihelion.

\item The heliographic latitude appearing at the centre of each solar map
oscillates annually due to the Earth's orbital inclination. Thus the number of
samples taken at each heliographic latitude varies between maps, such that
observational altitude is not independent of the heliographic latitude. Within
a map there will be more samples along the visible equator, and on average
along the heliographic equator, where the disc is widest than towards the
poles. During winter, the observation altitude is very low,
$\sim6^{\circ}$, which adds a bias if favour of one or other
hemisphere of the Sun. There is a risk that observed asymmetry may
in fact arise due to this bias.

\item There are around 100 times more maps recorded during summer than
during winter, due to differences in solar altitude and daylight hours.

\item The number of maps, the number of observation days per year, and the
number of observations per day vary greatly over the history of observations
(from 1978 to 2024, see also Fig.\,\ref{maps_per_month}).

\end{enumerate}


To correct for these biases in the data, we apply series of normalisations to weight the samples equally. This is to prevent ringing artifacts when fitting the polynomials for the limb model.


\begin{enumerate}
\item Samples $s_i$ for each map are weighted by ca. $r_i^{-1}$ as to have all radial bins an equal contribution.
\item The total weight of a map is independent of the amount of samples on the map.
\item Each observation day within a year has equal weight.
\item Each year from 1978 to 2021 has equal weight. For the incomplete years their weight is relative to the length of data during that year.
\item Each two-dimensional bin of observational sky altitude vs. heliographic latitude is given equal weight.
\end{enumerate}

In the first two stages, the weights are calculated so as to give each
radial distance an equal representation. Assuming the samples to be
uniformly distributed on the visible solar disk, the number of samples within a radial bin $r \pm
\delta r$ would be $\propto \, \delta r$, yielding a sample weight $n_S^{-1} \left( \operatorname{max}
\left\{ r_i, n_S^{-1/2} \right\} + r_i \right)^{-1}$. Here $n_S$ is the amount of samples on the map.
The max operator is used to clip the weight to avoid very large weights for the center samples.

The weighing scheme is not applied to the butterfly diagrams in Fig.~\ref{butterfly_clear_raw}. We have, however, tested relevant
combinations of different equalisations and the corresponding butterfly diagrams are available for
review. Different normalisation options do not significantly change the appearance of the butterfly diagrams.


\section{Chromosphere effects}\label{sect:chrom}

In addition to extracting and characterizing features from solar maps, we are
making parallel development with an inversion problem approach. Having obtained
a solar map from MRO, we describe a model for the chromosphere as well as the
MRO telescope which records the map. Each configuration of the model produces a
particular solar map, and we compute this map by ray-tracing and convolution.
We fine-tune the parameters of this model until the map obtained best fits the
actual observation.

As a first approximation, the chromosphere is a spherically symmetric shell
where the physical properties of the plasma depend only on radial distance $r$
from the center of the Sun. When we observe this object using the MRO
$\SI{14}{m}$ dish, we obtain a radially symmetric disk. In the middle of this
disk, we observe an intensity of one Quiet Sun Level. This is where the line of
sight from the observer makes a straight angle with the chromosphere. When
travelling towards the limb, we obtain a profile similar to what is seen in
Fig.\,\ref{limb_brightening}. Consider angle $\alpha$ between the shell and the
line of sight from the observer. The observed intensity on the map is a
function of this angle, namely, $I(\alpha)$. In the middle of the map, $\alpha
= 90^{\circ}$, whereas at the limb, $\alpha \to 0$.

Consider now a point $\bm{w}$ on a line of sight which originates from observer $\bm{o}$ at Earth and proceeds approximately towards the Sun in unit direction $\bm{\hat{b}}$:
\eqnl{line-of-sight}{
\bm{w}(a) = \bm{o} + \left( a + \s{a}{offset} \right) \bm{\hat{b}} \text{.}
}
We denote the travelled length with $a$, starting from offset $\s{a}{offset}$,
which is at the effective solar surface of radius $R_0$. This marks a
reasonable outer boundary for the chromosphere. We place the origin at the
center of the Sun and assume that the space is transparent outside radius $R_0$
and opaque inside radius $R_0 - \s{q}{max}$, thus setting the physical depth of
the chromosphere as $\s{q}{max}$. Function $q(a)$ maps the length travelled
along line of sight, measured as $a$, to the physical depth $q$ as:
\eqnl{line-of-sight-height}{
q(a) = R_0 - \norm{\bm{o} + \left( a + \s{a}{offset} \right) \bm{\hat{b}}} \text{.}
}
Thus within the chromosphere:
\eqnl{line-of-sight-length}{
0 \le a \le \s{a}{max} \; \text{and} \; 0 \le q \le \s{q}{max} \; \text{, with}
}
\eqnl{line-of-sight-length2}{
q(0) = 0 \; \text{and} \; q(\s{a}{max}) \in \{ 0, \s{q}{max} \} \text{.}
}
We emphasize that $\s{a}{offset}$ in Eq. \ref{line-of-sight} is a function
of $\bm{o}$ and $\bm{\hat{b}}$, chosen individually for each line to match the
chromosphere outer boundary. At the central region of the disk, $q(\s{a}{max})
= \s{q}{max}$ and the line of sight reaches the bottom shell, while at the limb
it is also possible that we see through the chromosphere and have
$q(\s{a}{max}) = 0$. For now, we do not consider this effect.

We define two functions for describing the chromosphere, $\eta(q)$ and $T(q)$.
These are for opacity and signal temperature, respectively, as measured for
$\SI{37}{GHz}$ radiation as a function of depth, $q$. Between the two reference
spheres, where $0 < q < \s{q}{max}$, the opacity takes finite non-zero values.
Otherwise
\eqnl{line-of-sight-opacity}{
\eta(q) &= 0      &\; \text{when} \; q \le 0 \; \text{and} \\
\eta(q) &= \infty &\; \text{when} \; q \ge \s{q}{max} \text{.}
}
The signal temperature $T(q)$ is also measured in $q \in [0, \s{q}{max}]$, and we denote the bottom temperature as $\s{T}{bottom} = T(\s{q}{max})$.

Consider a beam of radiation with intensity $I(a)$ travelling along the line of
sight (Eq. \ref{line-of-sight}) towards the observer at $\bm{o}$. At
position measured with coordinate $a$, the radiation has recently interacted
with the plasma along an infinitesimal path length $\dd a$. The radiation at
$a$ is thus a sum of that emitted between positions $a+\dd a$ and $a$ and what
came at $a +\dd a$ and has not been absorbed until $a$:
\eqnl{line-of-sight-interaction}{
I(a) = I(a + \dd a) * (1 - \eta(q(a)) \dd a) + T(q) \eta(q(a)) \dd a \text{.}
}
There is an infinitesimal change in intensity, $\dd I = I(a + \dd a) - I(a)$, which becomes:
\eqnl{line-of-sight-interaction2}{
\dd I &=& \left( I(a + \dd a) - T(q) \right) \eta(q(a)) \dd a \\
&\approx& \left( I(a) - T(q) \right) \eta(q(a)) \dd a \text{.}
}
The opacity $\eta$ thus defines the strength of the coupling between local
plasma and the beam. Since Eq.\,\ref{line-of-sight-interaction2} is linear for
$I$, we can obtain the solution $I(a)$ as sum of solutions for arbitrary
temperature profiles, as long as these profiles add up as $T\left(q(a)\right)$.
We use Dirac delta functions for this and define a particular element
$T_{a^{\prime}}(a)$ as:
\eqnl{line-of-sight-profile}{
T_{a^{\prime}}(a) = T\left(q(a^{\prime})\right) \delta(a - a^{\prime}) \text{.}
}
The opaque background at $a = \s{a}{max}$ with temperature $T(\s{a}{max})$ can be treated as an individual contribution to the observed intensity $I(0)$ as:
\eqnl{opacity_contribution}{
I(0) = \s{I}{finite}(0) + \s{I}{opaque}(0) \text{.}
}
Another option is to let the opacity asymptotically reach infinity as:
\eqnl{opacity_asymptotic}{
\eta(\s{q}{max} - \epsilon) \propto \epsilon^{-n} \; \text{with} \; n \ge 1 \text{,}
}
so that the spatial integral $\intef{q}{0}{\s{q}{max}}{\eta(q)}$ becomes infinite.

The temperature profile $T\left(q(a)\right)$ can be reconstructed as an integral of contributions from different depth:
\eqnl{line-of-sight-profile2}{
T\left(q(a)\right) = \intef{a^{\prime}}{0}{\s{a}{max}}{T_{a^{\prime}}(a)} \text{.}
}
We denote the corresponding solution as $I_{a^{\prime}}$, which satisfies:
\eqnl{line-of-sight-profile3}{
\dd I_{a^{\prime}} = \left( I_{a^{\prime}}(a) - T_{a^{\prime}}(a) \right) \eta(q(a)) \dd a \text{.}
}
On the region where $a \ne a^{\prime}$ and thus $T_{a^{\prime}}(a) = 0$, we can integrate Eq.\,\ref{line-of-sight-profile3} as:
\eqnl{line-of-sight-profile4}{
\int \frac{\dd I_{a^{\prime}}}{I_{a^{\prime}}} = \ln \left( I_{a^{\prime}} \right) = \int \eta \left(q(a)\right) \dd a \text{.}
}
Thus for small $\epsilon$:
\eqnl{line-of-sight-profile5}{
\ln \left( I_{a^{\prime}}(a^{\prime}-\epsilon) \right) - \ln \left( I_{a^{\prime}}(0) \right) = \intef{a}{0}{a^{\prime}-\epsilon}{\eta\left(q(a)\right)} \text{.}
}
This results in:
\eqnl{line-of-sight-profile6}{
I_{a^{\prime}}(0) = I_{a^{\prime}}(a^{\prime}-\epsilon) * \exp \left( -\intef{a}{0}{a^{\prime}-\epsilon}{\eta\left(q(a)\right)} \right) \text{.}
}
We set the boundary condition as $I_{a^{\prime}}(\s{a}{max}) = 0$, so that also:
\eqnl{line-of-sight-profile7}{
I_{a^{\prime}}(a^{\prime}+\epsilon) = 0 \text{.}
}
To handle the discontinuity of $I_{a^{\prime}}(a)$ at $a = a^{\prime}$, we integrate over infinitesimal length $[a^{\prime} - \epsilon, a^{\prime} + \epsilon]$ to get:
\eqnl{line-of-sight-profile8}{
I_{a^{\prime}}(a^{\prime}-\epsilon) \approx \intef{a}{a^{\prime}-\epsilon}{a^{\prime}+\epsilon}{T_{a^{\prime}}(a) \eta(q(a))} = \eta(q(a^{\prime})) T(a^{\prime}) \text{.}
}
We substitute this into Eq.\,\ref{line-of-sight-profile6} and ignore the infinitesimal gap $\epsilon$:
\eqnl{line-of-sight-profile9}{
I_{a^{\prime}}(0) = \eta(q(a^{\prime})) T(a^{\prime}) * \exp \left( -\intef{a}{0}{a^{\prime}}{\eta\left(q(a)\right)} \right) \text{.}
}
We integrate over all temperature profiles to get the contribution from finite opacity region $q \in [0, \s{q}{max})$:
\eqnl{line-of-sight-observed}{
\s{I}{finite}(0) = \intef{a^{\prime}}{0}{\s{a}{max}}{I_{a^{\prime}}(0)} \\
= \intef{a^{\prime}}{0}{\s{a}{max}}{\eta(q(a^{\prime})) T(q(a^{\prime})) * \exp \left( -\intef{a}{0}{a^{\prime}}{\eta\left(q(a)\right)} \right)} \text{.} \label{line-of-sight-observed2}
}
For the opaque background contribution in Eq. \ref{opacity_contribution}, we have:
\eqnl{opacity-contribution2}{
\s{I}{opaque}(0) = T(\s{q}{max}) * \exp \left( -\intef{a}{0}{\s{a}{max}}{\eta \left( q(a) \right)} \right) \text{.}
}
Instead of using physical distances $a$ and $a^{\prime}$, we'll switch to using optical distance $z$ and $z^{\prime}$ as follows:
\eqnl{optical-distance}{
\dd z = \eta(q(a)) \dd a \; \text{and} \; \dd z^{\prime} = \eta(q(a^{\prime})) \dd a^{\prime} \text{.}
}
The scales of $a^{\prime}$ and $z$ will share the same origin, so that:
\eqnl{optical-distance2}{
z(a) = \intef{a^{\prime}}{0}{a}{\eta(q(a^{\prime}))} \text{.}
}
We may further approximate that the chromosphere is spatially a thin spherical shell, so that the observed temperature
$I(0)$ is determined only by $\alpha$, the angle between line of sight and the shell. Thus the dependency between $q$
and $a^{\prime}$ is linear:
\eqnl{chromo-linear-approx}{
q(a^{\prime}) = a^{\prime} * \sin \alpha \text{.}
}
We use symbol $z_{\perp}$ as coordinate for the perpendicular case where $\alpha = 90^{\circ}$. The optical thickness of the chromoshere, seen perpendically, is $\s{z}{max}$:
\eqnl{optical-thickness-perp}{
\s{z}{max} &=& \intef{q}{0}{\s{q}{max}}{\eta(q)} \text{.}
}
We have for general $\alpha$:
\eqnl{optical-angle}{
z_{\perp}(a^{\prime}) &=& z(a^{\prime}) * \sin \alpha = \sin \alpha \intef{a}{0}{a^{\prime}}{\eta(a)} \; \text{.}
}
We substitute Eq.s \ref{optical-distance} \textemdash \ref{optical-angle} to Eq.s \ref{line-of-sight-observed} \textemdash \ref{opacity-contribution2}:
\eqnl{intensity-finite}{
\s{I}{finite}(0) = \frac{1}{\sin \alpha} \intef{z_{\perp}}{0}{\s{z}{max}}{T(z_{\perp}) * \exp \left( -\frac{z_{\perp}}{\sin \alpha} \right)} \; \text{and}
}
\eqnl{intensity-opague}{
\s{I}{opaque}(0) = T(\s{z}{max}) * \exp \left( -\frac{\s{z}{max}}{\sin \alpha} \right) \text{.}
}

When the angle of attack, $\alpha$, is very small, we ultimately observe the temperature at the very top of chromosphere:
\eqnl{opacity8}{
\alpha \to 0 \implies I(0) \to T(0) \text{.}
}
Thus when probing the solar limb, we are measuring the temperature of higher altitudes of the chromosphere.

Different temperature distributions as $T(z_{\perp})$ produce different limb
brightening patterns when convoluted with the antenna beam 
We have
tried various polynomial functions, truncated to a variable temperature value
at the opaque background at a variable optical depth $\s{z_{\perp}}{max}$. The
parameters of these functions are optimised using least squares methods, where
we minimize the difference between modellen and observed solar maps. This
requires the antenna beam to be parametrised as well as the physical size of
the Sun. The size of the Sun is optimal when the corresponding temperature
profile is as simple as possible e.g. has no extrema in $z_{\perp} \in (0,
\s{z_{\perp}}{max})$. If we use too large value for the solar radius, the
optimised temperature profile will dip at $z_{\perp} = 0$ as it tries to
compensate for the background sky, where as too small radius will result in
having a mid-depth temperature minimum when the model compensates for the
oversised rim.

Point sources, such as observed active regions, can be modelled as delta peaks
of intensity, which are smeared out by the antenna convolution. However, since
the emission might be of gyromagnetic origin, it does not radiate
isotropically. This phenomenon has to be treated somehow in future work.
\end{appendix}
\end{document}